\documentclass[preprint]{ptephy_v1}%%%%%% to generate preprint number

\preprintnumber{KYUSHU-HET-286} %%% %%% Insert preprint number here

\usepackage{graphics}

\usepackage{amsmath} %for dealing with mathematics,
\usepackage{amsthm} %for dealing with theorem environments,
\usepackage{url} %It provides better support for handling and breaking URLs.
\usepackage{stmaryrd}
\usepackage{tikz}
\usepackage{physics}
\usepackage{xcolor}

\numberwithin{equation}{section}

\allowdisplaybreaks

\begin{document}

\title{Action of the axial $U(1)$ non-invertible symmetry on the
't~Hooft line operator: A lattice gauge theory study}

%%%% To generate auto affiliation numbers please use \author{}\affil{} command

\author[1]{Yamato Honda}
\affil[1]{Department of Physics, Kyushu University, 744 Motooka, Nishi-ku,
Fukuoka 819-0395, Japan}

\author[1]{Soma Onoda}

\author[1]{Hiroshi Suzuki}

%% \author{Insert second author name here}
%% \affil{Insert second author address here}

%% \author{Insert third author name here}
%% \author[3]{Insert fourth author name here} %%% Use optional bracket [3] to change the respective address
%% \affil{Insert third author address here}

%% \author{Insert last author name here\thanks{These authors contributed equally to this work}}
%% \affil{Insert last author address here}

%%% To include the collaborator name... Please use the command "\collaborator"
%%% For example: \collaborator{ATLAS Collaboration}

\begin{abstract}%
We study how the symmetry operator of the axial $U(1)$ non-invertible symmetry
acts on the 't~Hooft line operator in the $U(1)$ gauge theory by employing
the modified Villain-type lattice formulation. We model the axial anomaly by a
compact scalar boson, the ``QED axion''. For the gauge invariance, the simple
't~Hooft line operator, which is defined by a line integral of the dual $U(1)$
gauge potential, must be ``dressed'' by the scalar and $U(1)$ gauge fields. A
careful consideration on the basis of the anomalous Ward--Takahashi identity
containing the 't~Hooft operator with the dressing factor and a precise
definition of the symmetry operator on the lattice shows that the symmetry
operator leaves no effect when it sweeps out a 't~Hooft loop operator. This
result appears inequivalent with the phenomenon concluded in the continuum
theory. In an appendix, we demonstrate that the half-space gauging of the
magnetic $\mathbb{Z}_N$ 1-form symmetry, when formulated in an appropriate
lattice framework, leads to the same conclusion as above. A similar result is
obtained for the axion string operator.
\end{abstract}

\subjectindex{B01,B04,B31}

\maketitle

\section{Introduction and summary}
\label{sec:1}
The idea that the existence of quantum anomaly does not necessarily implies the
absence of symmetries~\cite{Choi:2022jqy,Cordova:2022ieu} from a modern
perspective on the symmetry~\cite{Gaiotto:2014kfa} (see
Refs.~\cite{Schafer-Nameki:2023jdn,Bhardwaj:2023kri,Shao:2023gho} for reviews)
is quite interesting and should be further studied from various points of view.
In~Refs.~\cite{Choi:2022jqy,Cordova:2022ieu}, it is pointed out that, in $U(1)$
gauge theory, one can construct a topological gauge invariant symmetry operator
which generates the axial rotation with discrete angles on the Dirac fermion,
despite the axial $U(1)$ anomaly. This can be regarded as the existence of a
symmetry from the modern perspective, but the symmetry operator generates the
non-invertible symmetry, which is a subject of vigorous recent
studies~\cite{Aasen:2016dop,Bhardwaj:2017xup,Chang:2018iay,Thorngren:2019iar,Komargodski:2020mxz,Koide:2021zxj,Choi:2021kmx,Kaidi:2021xfk,Hayashi:2022fkw,Choi:2022zal,Kaidi:2022uux,Roumpedakis:2022aik,Bhardwaj:2022yxj,Cordova:2022ieu,Choi:2022jqy,Bhardwaj:2022lsg,Karasik:2022kkq,GarciaEtxebarria:2022jky,Choi:2022fgx,Yokokura:2022alv,Nagoya:2023zky,Anber:2023mlc}.

One of the interesting questions considered
in~Refs.~\cite{Choi:2022jqy,Cordova:2022ieu} is that how the symmetry operator
acts on the 't~Hooft line operator~\cite{tHooft:1977nqb,Yoneya:1978dt}.
In~Refs.~\cite{Choi:2022jqy,Cordova:2022ieu}, it is concluded that, when the
symmetry operator sweeps out a 't~Hooft line operator along a loop~$\gamma$, it
acquires a surface operator of the $U(1)$ field strength whose boundary
is~$\gamma$. Although this phenomenon might be understood as the Witten
effect~\cite{Witten:1979ey} on the 't~Hooft line (i.e., the worldline of the
monopole) in the presence of the axial
anomaly~$\epsilon_{\mu\nu\rho\sigma}f_{\mu\nu}(x)f_{\rho\sigma}(x)$, it is somewhat
puzzling because, at least naively, the 't~Hooft line operator does not receive
the axial transformation.

Let us take the anomalous Ward--Takahashi (WT) identity in the $U(1)$ gauge
theory:
\begin{equation}
   \left\langle
   \exp\left\{
   \frac{i\alpha}{2}\int_{\mathcal{V}_4}d^4x\,
   \left[\partial_\mu j_{5\mu}(x)
   -\frac{\mathrm{e}^2}{16\pi^2}
   \epsilon_{\mu\nu\rho\sigma}f_{\mu\nu}(x)f_{\rho\sigma}(x)
   \right]\right\}
   \mathcal{O}
   \right\rangle
   =\left\langle\mathcal{O}^\alpha\right\rangle,
\label{eq:(1.1)}
\end{equation}
where $\alpha$ is the axial rotation angle, $j_{5\mu}(x)$ is the Noether
current associated with the axial $U(1)$ transformation,
$\mathrm{e}\in\mathbb{Z}$ is the $U(1)$ charge of the Dirac fermion, and
$f_{\mu\nu}(x)$ is the field strength of the $U(1)$ gauge field. The
superscript~$\alpha$ on the right-hand side implies that the
observable~$\mathcal{O}$ within the 4-volume~$\mathcal{V}_4$ is axially
rotated; $\psi(x)^\alpha=e^{i(\alpha/2)\gamma_5}\psi(x)$
and~$\Bar{\psi}(x)^\alpha=\Bar{\psi}(x)e^{i(\alpha/2)\gamma_5}$. This identity can
be derived as usual by applying the axial rotation on the Dirac fermion and
considering the resulting axial anomaly. The symmetry operator
in~Refs.~\cite{Choi:2022jqy,Cordova:2022ieu} precisely generates the
transformation in~Eq.~\eqref{eq:(1.1)} with~$\alpha=2\pi p/N$ (here,
$p$ and~$N$ are coprime integers). However, since the
identity~\eqref{eq:(1.1)} is obtained by the axial rotation on the matter
field, i.e., the fermion, it is not intuitively obvious why the 't~Hooft
operator has a non-trivial effect in the identity~\eqref{eq:(1.1)}.

In this paper, we study this question by employing the lattice gauge theory
which provides an unambiguous definition of gauge theory. This problem is
challenging from the lattice gauge theory side, because it is difficult
to introduce the 't~Hooft line operator in lattice gauge theory with matter
fields with the axial anomaly or with the topological $\theta$ term. So far,
in lattice gauge theory, we have a control over the structure of the axial
anomaly of the fermion and the topological charge in non-Abelian gauge theory
only under the admissibility
condition~\cite{Luscher:1981zq,Hernandez:1998et,Luscher:1998kn}. A
consideration regarding the axial $U(1)$ non-invertible
symmetry~\cite{Choi:2022jqy,Cordova:2022ieu} on the basis of this control is
given in~Ref.~\cite{Honda:2024yte}. However, the admissibility implies that a
vanishing monopole current (i.e., the Bianchi identity) and thus it prohibits
the 't~Hooft line operator. A possible way out from this dilemma is to
``excise'' lattice sites along the monopole worldline and this method actually
works for a 2D compact scalar field on the
lattice~\cite{Abe:2023uan,Morikawa:2024zyd}. This ``excision method'' however
creates boundaries on the lattice and at the moment it is difficult to say
something how the axial anomaly is affected by the presence of such lattice
boundaries (see Ref.~\cite{Luscher:2006df} for a related study).

In the present paper, to study the action of the non-invertible symmetry
operator on the 't~Hooft line operator while avoiding the above
difficulties in lattice gauge theory, we model the axial anomaly by using a
$2\pi$~periodic real scalar field. The continuum Euclidean action is (here,
$\mu$ is a constant of the mass dimension~1),
\begin{equation}
   \int d^4x\,\left[
   \frac{\mu^2}{2}\partial_\mu\phi(x)\partial_\mu\phi(x)
   +\frac{i\mathrm{e}^2}{32\pi^2}\phi(x)
   \epsilon_{\mu\nu\rho\sigma}f_{\mu\nu}(x)f_{\rho\sigma}(x)
   \right],
\label{eq:(1.2)}
\end{equation}
for which the axial $U(1)$ transformation is defined by
\begin{equation}
   \phi(x)\to\phi(x)-\alpha,
\label{eq:(1.3)}
\end{equation}
and the Noether current is given by
\begin{equation}
   j_{5\mu}(x)=-2i\mu^2\partial_\mu\phi(x).
\label{eq:(1.4)}
\end{equation}
This can be regarded as a chiral Lagrangian for the $U(1)_A$ symmetry with
the Wess--Zumino term or the system of the
``QED axion''~\cite{Cordova:2022ieu,Hidaka:2020izy}.

On the other hand, to introduce the 't~Hooft line operator in $U(1)$ lattice
gauge theory, we employ the modified Villain-type lattice formulation developed
in~Ref.~\cite{Sulejmanpasic:2019ytl}. See also
Refs.~\cite{Gorantla:2021svj,Jacobson:2023cmr}. This lattice formulation,
containing the dual $U(1)$ gauge field from the beginning, allows a
straightforward introduction of the 't~Hooft line operator. For the gauge
invariance, however, we find that the simple 't~Hooft line operator, which is
defined by a line integral of the dual $U(1)$ gauge potential, must be
``dressed'' by the scalar and $U(1)$ gauge fields.\footnote{The necessity of
this dressing of the 't~Hooft line operator is discussed from a very general
grounds of a 5D TQFT in~Ref.~\cite{Antinucci:2024zjp}. See also
Ref.~\cite{Benedetti:2023owa} for a related analysis.} In the presence of the
't~Hooft line operator, we also need a modification of the lattice action for
the gauge invariance.\footnote{This modification of the lattice action itself,
guided by the construction of the $\theta$~term in the modified Villain
formulation~\cite{Jacobson:2023cmr}, appears highly non-trivial.} One can then
write down the anomalous WT identity containing the 't~Hooft line operator with
the dressing factor on the lattice. The anomalous WT identity becomes
non-trivial because of the dressing factor. A careful consideration of the
anomalous WT identity in terms of the symmetry operator of the non-invertible
symmetry on the lattice~\cite{Honda:2024yte} shows however that the symmetry
operator leaves no effect when it sweeps out a 't~Hooft loop operator. This
result, while it is consistent with the naive intuition, appears inequivalent
with the phenomenon concluded in~Refs.~\cite{Choi:2022jqy,Cordova:2022ieu}.
In~Appendix~\ref{sec:A}, we demonstrate that the half-space gauging of the
magnetic $\mathbb{Z}_N$ 1-form symmetry, which is employed
in~Refs.~\cite{Choi:2022jqy,Cordova:2022ieu} to study the action of the
symmetry operator on the 't~Hooft line operator, when formulated in an
appropriate lattice framework, leads to the same conclusion as that in the main
text. A similar analysis for the axion string operator~\cite{Hidaka:2020izy}
shows that it is also not affected from the symmetry operator.

The extension of our analysis to the system containing fermions and to the
system with the non-Abelian gauge symmetry is highly desirable, although we do
not know how to do that at the moment.

\section{Modified Villain-type lattice formulation}
\label{sec:2}
\subsection{$U(1)$ gauge theory}
\label{sec:2.1}
Our lattice gauge theory is defined on a finite hypercubic lattice of
size~$L$:\footnote{The Lorentz index is denoted by Greek letters $\mu$, $\nu$,
\dots, and runs over 0, 1, 2, and~3.}
\begin{equation}
   \Gamma
   :=\left\{x\in\mathbb{Z}^4\mathrel{}\middle|\mathrel{}0\leq x_\mu<L\right\}
\label{eq:(2.1)}
\end{equation}
and we assume periodic boundary conditions for all lattice fields.

Following the lattice formulation developed
in~Refs.~\cite{Sulejmanpasic:2019ytl,Jacobson:2023cmr}, as the lattice action
for the pure $U(1)$ gauge theory, we adopt\footnote{$\Hat{\mu}$ denotes the
unit vector in $\mu$~direction. $\partial_\mu$ is the forward difference
operator, $\partial_\mu f(x):=f(x+\Hat{\mu})-f(x)$.}
\begin{equation}
   \frac{1}{4g_0^2}\sum_{x\in\Gamma}f_{\mu\nu}(x)f_{\mu\nu}(x)
   +\frac{i}{2}\sum_{x\in\Gamma}
   \epsilon_{\mu\nu\rho\sigma}\Tilde{a}_\mu(x)
   \partial_\nu z_{\rho\sigma}(x+\Hat{\mu}).
\label{eq:(2.2)}
\end{equation}
In this modified Villain-type lattice formulation, the degrees of freedom of
the $U(1)$ gauge field are represented by real non-compact link
variables~$a_\mu(x)\in\mathbb{R}$ and integer plaquette
variables~$z_{\mu\nu}(x)=-z_{\nu\mu}(x)\in\mathbb{Z}$. The $U(1)$ field strength
in~Eq.~\eqref{eq:(2.2)} is defined from these variables by
\begin{equation}
   f_{\mu\nu}(x)=\partial_\mu a_\nu(x)-\partial_\nu a_\mu(x)+2\pi z_{\mu\nu}(x).
\label{eq:(2.3)}
\end{equation}
The basic principle of the formulation~\cite{Sulejmanpasic:2019ytl} is that the
lattice action and observables are required to be invariant under the
1-form $\mathbb{Z}$ gauge transformation,
\begin{equation}
   a_\mu(x)\to a_\mu(x)+2\pi m_\mu(x),\qquad
   z_{\mu\nu}(x)\to z_{\mu\nu}(x)-\partial_\mu m_\nu(x)+\partial_\nu m_\mu(x),
\label{eq:(2.4)}
\end{equation}
where $m_\mu(x)\in\mathbb{Z}$. $z_{\mu\nu}(x)$ is thus a 2-form $\mathbb{Z}$
gauge field. Then, one may take a gauge such that $-\pi<a_\mu(x)\leq\pi$
and~$z_{\mu\nu}(x)\in\mathbb{Z}$, which would be almost equivalent to a compact
$U(1)$ lattice gauge formulation which well captures topological properties in
continuum theory. The field strength~\eqref{eq:(2.3)} is invariant under the
1-form $\mathbb{Z}$ transformation~\eqref{eq:(2.4)} and the lattice
action~\eqref{eq:(2.2)} is thus consistent with this 1-form gauge invariance.

Since the field strength~\eqref{eq:(2.3)} is also invariant under
\begin{equation}
   a_\mu(x)\to a_\mu(x)+\partial_\mu\lambda(x),\qquad
   z_{\mu\nu}(x)\to z_{\mu\nu}(x),
\label{eq:(2.5)}
\end{equation}
where $\lambda(x)\in\mathbb{R}$, the lattice action~\eqref{eq:(2.2)} possesses
this ordinary 0-form $\mathbb{R}$ gauge invariance.

The field $\Tilde{a}_\mu(x)\in\mathbb{R}$ in~Eq.~\eqref{eq:(2.2)} is an
auxiliary field. Since $\partial_\nu z_{\rho\sigma}(x)$ in~Eq.~\eqref{eq:(2.2)}
are integers, there is another 1-form $\mathbb{Z}$ gauge symmetry, under which
\begin{equation}
   \Tilde{a}_\mu(x)\to\Tilde{a}_\mu(x)+2\pi\mathbb{Z}.
\label{eq:(2.6)}
\end{equation}
Then, by using this gauge symmetry, we may restrict
$-\pi<\Tilde{a}_\mu(x)\leq\pi$. Then, $\Tilde{a}_\mu(x)$ acts as a Lagrange
multiplier whose functional integral imposes the Bianchi identity,
\begin{equation}
   \epsilon_{\mu\nu\rho\sigma}
   \partial_\nu f_{\rho\sigma}(x)
   =2\pi\epsilon_{\mu\nu\rho\sigma}
   \partial_\nu z_{\rho\sigma}(x)
   =0.
\label{eq:(2.7)}
\end{equation}

Now, one of the advantages of the present lattice formulation is that the
auxiliary field~$\Tilde{a}_\mu(\Tilde{x})$ provides the dual $U(1)$ gauge
field. Using this, one may readily introduce the 't~Hooft line operator as
\begin{equation}
   T_{\mathrm{q}}(\gamma)\sim\exp\left[
   i\mathrm{q}\sum_{(x,\mu)\in\gamma}\Tilde{a}_\mu(x)\right],
\label{eq:(2.8)}
\end{equation}
where $\gamma$ is a curve on~$\Gamma$ and the link sum is taken along this
curve. The magnetic charge~$\mathrm{q}$ must be an integer for the invariance
under~Eq.~\eqref{eq:(2.6)}. When $T_\mathrm{q}(\gamma)$ is inserted in the
functional integral, the integral over the auxiliary field~$\Tilde{a}_\mu(x)$
enforces the breaking of the Bianchi identity along the curve~$\gamma$,
$(1/2)\epsilon_{\mu\nu\rho\sigma}\partial_\nu z_{\rho\sigma}(x)=\mathrm{q}$. This
represents the monopole worldline in the $\mu$~direction, i.e., the 't~Hooft
line in the $\mu$~direction.

\subsection{Gauge invariant lattice action of the periodic scalar field}
\label{sec:2.2}
Next, we introduce a $2\pi$~periodic scalar field~$\phi(x)$ as a simple system
which exhibits the axial $U(1)$ anomaly. For the kinetic term, we adopt (a 2D
analogue can be found in~Ref.~\cite{Berkowitz:2023pnz}, although we put the
scalar field on sites of the original lattice~$\Gamma$ instead of the dual
lattice)
\begin{align}
   \sum_{x\in\Gamma}\left[
   \frac{\mu^2}{2}\partial\phi(x,\mu)\partial\phi(x,\mu)
   +\frac{i}{2}\epsilon_{\mu\nu\rho\sigma}\partial_\mu\ell_\nu(x)
   \chi_{\rho\sigma}(x+\Hat{\mu}+\Hat{\nu})
   \right],
\label{eq:(2.9)}
\end{align}
where
\begin{equation}
   \partial\phi(x,\mu)
   :=\partial_\mu\phi(x)+2\pi\ell_\mu(x),
\label{eq:(2.10)}
\end{equation}
and $\ell_\mu(x)\in\mathbb{Z}$. In this modified Villain-type formulation, the
$2\pi$~periodic scalar field is represented by variables,
$\phi(x)\in\mathbb{R}$ and~$\ell_\mu(x)\in\mathbb{Z}$. Then, requiring the
invariance under the 0-form $\mathbb{Z}$ gauge transformation,
\begin{equation}
   \phi(x)\to\phi(x)+2\pi k(x),\qquad
   \ell_\mu(x)\to\ell_\mu(x)-\partial_\mu k(x),
\label{eq:(2.11)}
\end{equation}
where $k(x)\in\mathbb{Z}$, we may take a gauge such that $-\pi<\phi(x)\leq\pi$
and~$\ell_\mu(x)\in\mathbb{Z}$. The lattice action~\eqref{eq:(2.9)} is
consistent with this gauge invariance.

The field $\chi_{\mu\nu}(x)\in\mathbb{R}$ in~Eq.~\eqref{eq:(2.9)} is an another
auxiliary field. By using the gauge invariance under
\begin{equation}
   \chi_{\mu\nu}(x)\to\chi_{\mu\nu}(x)+2\pi\mathbb{Z},
\label{eq:(2.12)}
\end{equation}
we can restrict $-\pi<\chi_{\mu\nu}(x)\leq\pi$. The functional integral
over~$\chi_{\mu\nu}(x)$ then imposes the Bianchi identity for the scalar field,
\begin{equation}
   \epsilon_{\mu\nu\rho\sigma}\partial_\rho\partial\phi(x,\sigma)
   =2\pi\epsilon_{\mu\nu\rho\sigma}\partial_\rho\ell_\sigma(x)=0.
\label{eq:(2.13)}
\end{equation}

To the free part~\eqref{eq:(2.9)}, we want to further add a lattice
counterpart of~$\phi\epsilon_{\mu\nu\rho\sigma}f_{\mu\nu}f_{\rho\sigma}$, which
generates the axial anomaly under the shift~$\phi\to\phi+\alpha$. To find and
write down desired expressions, the notion of the cochain, coboundary
operator and the cup products on the hypercubic
lattice~\cite{Chen:2021ppt,Jacobson:2023cmr} is quite helpful. Thus, we
identify lattice fields, $a_\mu(x)$, $z_{\mu\nu}(x)$, $f_{\mu\nu}(x)$,
$\Tilde{a}(x)$, $\phi(x)$, $\ell_\mu(x)$, $\partial\phi(x,\mu)$, $\chi(x)$,
with cochains, $a$, $z$, $f$, $\Tilde{a}$, $\phi$, $\ell$, $\partial\phi$,
$\chi$ with obvious ranks, respectively. In particular, corresponding
to~Eqs.~\eqref{eq:(2.3)} and~\eqref{eq:(2.10)}, we have
\begin{equation}
   f=\delta a+2\pi z,\qquad\partial\phi=\delta\phi+2\pi\ell,
\label{eq:(2.14)}
\end{equation}
where $\delta$ is the coboundary operator~\cite{Chen:2021ppt,Jacobson:2023cmr}.
The coboundary operator is nilpotent~$\delta^2=0$. Also, the Bianchi identities,
Eqs.~\eqref{eq:(2.7)} and~\eqref{eq:(2.13)}, are simply expressed as
\begin{equation}
   \delta f=2\pi\delta z=0,\qquad
   \delta\partial\phi=2\pi\delta l=0.
\label{eq:(2.15)}
\end{equation}

Now, as a lattice counterpart
of~$\phi\epsilon_{\mu\nu\rho\sigma}f_{\mu\nu}f_{\rho\sigma}$, we start with the
expression
\begin{align}
   \sum_{\text{hypercube}\in\Gamma}
   \phi\cup f\cup f
   =\sum_{x\in\Gamma}
   \frac{1}{4}\phi(x)\epsilon_{\mu\nu\rho\sigma}
   f_{\mu\nu}(x)f_{\rho\sigma}(x+\Hat{\mu}+\Hat{\nu}).
\label{eq:(2.16)}
\end{align}
The particular way of shifts of lattice sites on the right-hand side
corresponds to the cup product~\cite{Chen:2021ppt,Jacobson:2023cmr} on the
left-hand side. This particular shift structure is well-known in the context of
the axial anomaly in lattice $U(1)$ gauge
theory~\cite{Luscher:1998kn,Fujiwara:1999fi,Fujiwara:1999fj}.
Under the cup product, the coboundary operator~$\delta$ satisfies the Leibniz
rule.\footnote{These notions have been also known as the non-commutative
differential calculus~\cite{Fujiwara:1999fi}.}

Starting from~Eq.~\eqref{eq:(2.16)}, we seek an appropriate combination which
is invariant under all the following gauge transformations: The 0-form
$\mathbb{Z}$ gauge transformation~\eqref{eq:(2.11)}
\begin{equation}
   \phi\to\phi+2\pi k,\qquad
   \ell\to\ell-\delta k,
\label{eq:(2.17)}
\end{equation}
where $k\in\mathbb{Z}$ is a 0-cochain, the 1-form $\mathbb{Z}$ gauge
transformation~\eqref{eq:(2.4)}
\begin{equation}
   a\to a+2\pi m,\qquad
   z\to z-\delta m,
\label{eq:(2.18)}
\end{equation}
where $m\in\mathbb{Z}$ is a 1-cochain and, finally, the 0-form $\mathbb{R}$
gauge transformation~\eqref{eq:(2.5)}
\begin{equation}
   a\to a+\delta\lambda,\qquad
   z\to z,
\label{eq:(2.19)}
\end{equation}
where $\lambda\in\mathbb{R}$ is a 0-cochain.

While referring the construction of the $\theta$~term
in~Ref.~\cite{Jacobson:2023cmr}, with some trial and error, we find that the
following combination possesses desired invariant properties:\footnote{%
In Eq.~\eqref{eq:(2.21)}, the first line can be regarded as a natural
lattice transcription of $\phi\wedge f\wedge f$, which is consistent with the
gauge symmetries (this line alone is gauge invariant if the Bianchi identities
hold). Although the second line is of a somewhat complicated composition, it is
proportional to~$\delta z$ and thus vanishes if the Bianchi
identity~$\delta z=0$ holds. When the Bianchi identity~$\delta z=0$ holds, the
explicit form of the lattice action is then
\begin{align}
   \frac{i\mathrm{e}^2}{8\pi^2}I&
   =\frac{i\mathrm{e}^2}{32\pi^2}\sum_{x\in\Gamma}
   \biggl[
   \phi(x)
   \epsilon_{\mu\nu\rho\sigma}f_{\mu\nu}(x)f_{\rho\sigma}(x+\Hat{\mu}+\Hat{\nu})
\notag\\
   &\qquad\qquad\qquad{}
   -4\pi\epsilon_{\mu\nu\rho\sigma}
   \ell_\mu(x)
   a_\nu(x+\Hat{\mu})f_{\rho\sigma}(x+\Hat{\mu}+\Hat{\nu})
\notag\\
   &\qquad\qquad\qquad{}
   -8\pi^2\epsilon_{\mu\nu\rho\sigma}
   \ell_\mu(x)
   z_{\nu\rho}(x+\Hat{\mu})
   a_\sigma(x+\Hat{\mu}+\Hat{\nu}+\Hat{\rho})
   \biggr].
\label{eq:(2.20)}
\end{align}
}
\begin{align}
   I&:=\sum_{\text{hypercube}\in\Gamma}
   \left(\phi\cup f\cup f-2\pi\ell\cup a\cup f
   -4\pi^2\ell\cup z\cup a\right)
\notag\\
   &\qquad{}
   +\sum_{\text{hypercube}\in\Gamma}
   \bigl\{
   \phi\cup\left[
   -2\pi a\cup\delta z+2\pi\delta z\cup a
   +2\pi\delta(a\cup_1\delta z)
   +4\pi^2z\cup_1\delta z\right]
\notag\\
   &\qquad\qquad\qquad\qquad{}
   -4\pi^2\ell\cup\left(a\cup_1\delta z\right)
   \bigr\},
\label{eq:(2.21)}
\end{align}
where $\cup_1$ is a higher cup product~\cite{Chen:2021ppt,Jacobson:2023cmr}.
We do not need the explicit form of~$\cup_1$ in what follows but what is
crucial to us is that $\cup_1$ enables one to exchange the ordering of the cup
product of two cochains (here $\alpha$ and~$\beta$ are $p$ and $q$ cochains,
respectively):
\begin{equation}
   (-1)^{pq}\beta\cup\alpha
   =\alpha\cup\beta+(-1)^{p+q}
   \left[\delta(\alpha\cup_1\beta)
   -\delta\alpha\cup_1\beta
   -(-1)^p\alpha\cup_1\delta\beta\right].
\label{eq:(2.22)}
\end{equation}
Using Eq.~\eqref{eq:(2.22)}, under~Eqs.~\eqref{eq:(2.17)}--\eqref{eq:(2.19)},
we see that the combination~\eqref{eq:(2.21)} changes into
\begin{align}
   I&\stackrel{\text{0-form $\mathbb{Z}$}}{\to}
   I+8\pi^3\mathbb{Z}
\notag\\
   &\stackrel{\text{1-form $\mathbb{Z}$}}{\to}
   I+\sum_{x\in\Gamma}
   \left(-8\pi^2\phi\cup m\cup\delta z+4\pi^2\delta\ell\cup m\cup a\right)
   +8\pi^3\mathbb{Z}
\notag\\
   &\stackrel{\text{0-form $\mathbb{R}$}}{\to}
   I+\sum_{x\in\Gamma}
   \bigl\{
   \left(-4\pi\phi\cup\delta\lambda+8\pi^2\ell\cup\lambda\right)\cup\delta z
\notag\\
   &\qquad\qquad\qquad\qquad{}
   +\delta\ell\cup\left[
   -2\pi\lambda\cup\delta a
   -4\pi^2\left(\lambda\cup z+z\cup\lambda+\lambda\cup_1\delta z\right)
   \right]
   \bigr\},
\label{eq:(2.23)}
\end{align}
where we have used $\delta^2z=0$, which always holds because the field $z$ is
single-valued on~$\Gamma$. Note that all the gauge breakings
in~Eq.~\eqref{eq:(2.23)} are proportional to the Bianchi
identities in~Eq.~\eqref{eq:(2.15)}. Therefore, setting the lattice action as
\begin{equation}
   S:=\sum_{\text{hypercube}\in\Gamma}\left(\frac{1}{2g_0^2}f\cup\star f
   +i\Tilde{a}\cup\delta z\right)
   +\sum_{\text{hypercube}\in\Gamma}
   \left(\frac{\mu^2}{2}
   \partial\phi\cup\star\partial\phi+i\delta\ell\cup\chi\right)
   +\frac{i\mathrm{e}^2}{8\pi^2}I,
\label{eq:(2.24)}
\end{equation}
where the first two terms are Eqs.~\eqref{eq:(2.2)} and~\eqref{eq:(2.9)},
respectively, by endowing the auxiliary fields with gauge transformations,
\begin{align}
   \Tilde{a}
   &\to\Tilde{a}+\mathrm{e}^2\phi\cup m
   +\frac{\mathrm{e}^2}{2\pi}(\phi\cup\delta\lambda-2\pi\ell\cup\lambda),
\notag\\
   \chi&\to
   \chi-\frac{\mathrm{e}^2}{2}m\cup a
   +\frac{\mathrm{e}^2}{4\pi}
   \left[\lambda\cup\delta a+2\pi\left(\lambda\cup z+z\cup\lambda
   +\lambda\cup_1\delta z\right)
   \right],
\label{eq:(2.25)}
\end{align}
the Boltzmann factor~$e^{-S}$ is invariant under all the gauge transformations
if $\mathrm{e}$ is an even integer.\footnote{Throughout this paper, we assume
this condition.}

In this way, we have obtained a gauge invariant lattice action and gauge
transformations of auxiliary fields. This knowledge then enables us to find a
gauge invariant 't~Hooft line operator. Note that the simple
definition~\eqref{eq:(2.8)} is not invariant under~Eq.~\eqref{eq:(2.25)}.
Therefore, the expectation value of~Eq.~\eqref{eq:(2.8)} is not sensible in
this lattice formulation with the scalar field. The operator must be
``dressed'' by the scalar and gauge fields for the gauge invariance. We find
that the following dressed one is a gauge invariant combination
when~$\delta\ell=0$:
\begin{equation}
   T_{\mathrm{q}}(\gamma)
   :=\exp\left\{
   i\mathrm{q}\left[
   \sum_{\text{link}\in\gamma}
   \left(\Tilde{a}-\frac{\mathrm{e}^2}{2\pi}\phi\cup a\right)
   +\sum_{\text{plaquette}\in\mathcal{R}}
   \frac{\mathrm{e}^2}{2\pi}\left(-2\pi\ell\cup a+\phi\cup f\right)
   \right]\right\},
\label{eq:(2.26)}
\end{equation}
where $\mathcal{R}$ is a 2D surface whose boundary is~$\gamma$,
$\partial(\mathcal{R})=\gamma$. If the present axion QED had the electric
1-form symmetry, then it would be possible to regard this operator as a
boundary of the topological symmetry operator of the 1-form symmetry. However,
the axion QED does not have such an electric 1-form symmetry because
$\delta\star f\neq0$. Corresponding to this fact, the surface~$\mathcal{R}$
attached to the 't~Hooft line~$\gamma$ is not topological. This is a property
in the continuum theory and is expected to survive under the (classical)
continuum limit.

Similarly, we can obtain a gauge invariant axion string operator (it is
basically given by a 2D surface integral of the auxiliary field~$\chi$) as
\begin{equation}
   S_{\mathrm{q}}(\sigma)
   :=\exp\left\{
   i\mathrm{q}\left[
   \sum_{\text{plaquette}\in\sigma}
   \left(\chi+\frac{\mathrm{e}^2}{4\pi}a\cup a\right)
   -\sum_{\text{cube}\in\mathcal{V}_3}
   \frac{\mathrm{e}^2}{4\pi}\left(f\cup a+2\pi a\cup z\right)
   \right]\right\},
\label{eq:(2.27)}
\end{equation}
where $\mathcal{V}_3$ is a 3-volume whose boundary is~$\sigma$,
$\partial(\mathcal{V}_3)=\sigma$. This is gauge invariant when~$\delta z=0$. We
use these expressions to investigate how these operators are affected by the 
symmetry operator of the axial $U(1)$ non-invertible symmetry.

\section{Action of the symmetry operator on the gauge invariant 't~Hooft line}
\label{sec:3}
Now, in the above lattice formulation, we derive an anomalous WT identity
containing the gauge-invariant 't~Hooft line operator~\eqref{eq:(2.26)}. We
first take the expectation value of the operators
\begin{equation}
   \left\langle T_{\mathrm{q}}(\gamma)\dotsb\right\rangle_{\mathrm{B}},
\label{eq:(3.1)}
\end{equation}
where the subscript~$\mathrm{B}$ stands for only the functional integral over
the scalar field sector ($\phi$ and~$\ell$) and the auxiliary
fields~($\Tilde{a}$ and~$\chi$) is carried out. We then consider the change of
integration variables of the form of a localized infinitesimal axial
transformation, $\phi(x)\to\phi(x)+\alpha(x)$ and~$\ell_\mu(x)\to\ell_\mu(x)$.
We assume that the abbreviated term~$\dotsb$ in~Eq.~\eqref{eq:(3.1)} is
invariant under this transformation. The lattice action~\eqref{eq:(2.24)} then
changes into (see also Eqs.~\eqref{eq:(2.9)} and~\eqref{eq:(2.20)})\footnote{%
Here, $\partial_\mu^*$ is the backward difference operator,
$\partial_\mu^*f(x):=f(x)-f(x-\Hat{\mu})$.}
\begin{align}
   &S-i\sum_{x\in\Gamma}\frac{\alpha(x)}{2}\left[
   \partial_\mu^*j_{5\mu}(x)
   -\frac{\mathrm{e}^2}{16\pi^2}\epsilon_{\mu\nu\rho\sigma}
   f_{\mu\nu}(x)f_{\rho\sigma}(x+\Hat{\mu}+\Hat{\nu})
   \right]
\notag\\
   &\qquad{}
   -\frac{i\mathrm{e}^2}{2\pi}\sum_{\text{hypercube}\in\Gamma}
   \alpha\cup\left(
   a\cup\delta z-\frac{1}{2}f\cup_1\delta z 
   \right),
\label{eq:(3.2)}
\end{align}
where the axial vector current is given by
\begin{equation}
   j_{5\mu}(x):=-2i\mu^2\partial\phi(x,\mu),
\label{eq:(3.3)}
\end{equation}
where $\partial\phi(x,\mu)$ is defined by~Eq.~\eqref{eq:(2.10)}; note that
this is invariant under the gauge transformation~\eqref{eq:(2.11)}.

On the other hand, owing to the dressing factor in~Eq.~\eqref{eq:(2.26)}, under
the infinitesimal shift, $\phi(x)\to\phi(x)+\alpha(x)$,
\begin{align}
   T_{\mathrm{q}}(\gamma)
   \to
   T_{\mathrm{q}}(\gamma)\exp\left[
   -\frac{i\mathrm{q}\mathrm{e}^2}{2\pi}\left(
   \sum_{\text{link}\in\gamma}
   \alpha\cup a
   -\sum_{\text{plaquette}\in\mathcal{R}}
   \alpha\cup f
   \right)\right].
\label{eq:(3.4)}
\end{align}
We then repeat this change of variables to make a finite axial
rotation, setting $\alpha(x)=\alpha$ within a 4-volume~$\mathcal{V}_4$,
$x\in\mathcal{V}_4\subset\Gamma$, and $\alpha(x)=0$ otherwise. We assume that
the curve~$\gamma$ and the surface~$\mathcal{R}$ are completely
within~$\mathcal{V}_4$. With the insertion of~Eq.~\eqref{eq:(2.26)}, the
integration over the auxiliary field~$\Tilde{a}$ enforces
$\delta z=\mathrm{q}\delta_3[\gamma]$, where $\delta_3[\gamma]$ is the delta
function 3-cochain along the curve~$\gamma$. If this is substituted
in~Eq.~\eqref{eq:(3.2)}, we see that the
$\sum_{\text{hypercube}\in\Gamma}\alpha\cup a\cup\delta z$ term
in~Eq.~\eqref{eq:(3.2)} and the $\sum_{\text{link}\in\gamma}\alpha\cup a$ term
in~Eq.~\eqref{eq:(3.4)} cancel each other out.

In this way, we have the identity,
\begin{align}
   &\left\langle
   \exp\left\{
   \frac{i\alpha}{2}\sum_{x\in\mathcal{V}_4}
   \left[\partial_\mu^*j_{5\mu}(x)
   -\frac{\mathrm{e}^2}{16\pi^2}\epsilon_{\mu\nu\rho\sigma}
   f_{\mu\nu}(x)f_{\rho\sigma}(x+\Hat{\mu}+\Hat{\nu})
   \right]\right\}T_{\mathrm{q}}(\gamma)\dotsb
   \right\rangle_{\mathrm{B}}
\notag\\
   &=
   \exp\left\{
   -\frac{i\mathrm{q}\mathrm{e}^2\alpha}{2\pi}
   \left[
   \sum_{\text{plaquette}\in\mathcal{R}}f
   -\sum_{\text{hypercube}\in\Gamma}
   \frac{1}{2}f\cup_1\delta_3[\gamma]
   \right]
   \right\}
   \left\langle
   T_{\mathrm{q}}(\gamma)\dotsb
   \right\rangle_{\mathrm{B}},
\label{eq:(3.5)}
\end{align}
or
\begin{align}
   &\left\langle
   \exp\left[
   \frac{i\alpha}{2}\sum_{\text{hypercube}\in\mathcal{V}_4}
   \left(\delta\star j_5
   -\frac{\mathrm{e}^2}{4\pi^2}f\cup f
   \right)\right]T_{\mathrm{q}}(\gamma)\dotsb
   \right\rangle_{\mathrm{B}}
\notag\\
   &=
   \exp\left\{
   -\frac{i\mathrm{q}\mathrm{e}^2\alpha}{2\pi}
   \left[
   \sum_{\text{plaquette}\in\mathcal{R}}f
   -\sum_{\text{hypercube}\in\Gamma}
   \frac{1}{2}f\cup_1\delta_3[\gamma]
   \right]
   \right\}
   \left\langle
   T_{\mathrm{q}}(\gamma)\dotsb
   \right\rangle_{\mathrm{B}}.
\label{eq:(3.6)}
\end{align}
This is the anomalous WT identity containing the gauge-invariant 't~Hooft line
operator.

Now, as shown in~Appendix~C of~Ref.~\cite{Honda:2024yte}, when
\begin{equation}
   \alpha=\frac{2\pi p}{N},
\label{eq:(3.7)}
\end{equation}
and $p\mathrm{e}^2$ and~$N$ are coprime integers, we can represent the symmetry
generator of the axial $U(1)$ non-invertible
symmetry~\cite{Choi:2022jqy,Cordova:2022ieu} for an arbitrary closed oriented
3-surface~$\mathcal{M}_3$ as\footnote{This was pointed out to us by Yuya
Tanizaki.}
\begin{equation}
   U_{2\pi p/N}(\mathcal{M}_3)
   =\exp\left\{
   \frac{i\pi p}{N}\sum_{\text{cube}\in\mathcal{M}_3}
   \left[\star j_5
   -\frac{\mathrm{e}^2}{4\pi^2}
   \left(a\cup f+2\pi z\cup a\right)
   \right]\right\}
   \mathcal{Z}_{\mathcal{M}_3}[z],
\label{eq:(3.8)}
\end{equation}
where $\mathcal{Z}_{\mathcal{M}_3}[z]$ is the partition function of a lattice 3D
TQFT (the level~$N$ BF theory; $s$ denotes the number of sites
of~$\Gamma$)~\cite{Honda:2024yte},
\begin{equation}
   \mathcal{Z}_{\mathcal{M}_3}[z]
   :=\frac{1}{N^s}
   \int\mathrm{D}[b]\mathrm{D}[c]\,
   \exp\left\{
   \frac{i\pi p\mathrm{e}^2}{N}
   \sum_{\text{cube}\in\mathcal{M}_3}
   \left[
   b\left(\delta c
   -z\right)
   -z\cup c
   \right]\right\},
\label{eq:(3.9)}
\end{equation}
where ($b_\mu(\Tilde{x})$ is residing on dual links)
\begin{equation}
   \mathrm{D}[b]:=
   \prod_{(x,\mu)\in\mathcal{M}_3}
   \left[\frac{1}{N}\sum_{b_\mu(\Tilde{x})=0}^{N-1}\right],\qquad
   \mathrm{D}[c]:=
   \prod_{(x,\mu)\in\mathcal{M}_3}\left[\sum_{c_\mu(x)=0}^{N-1}\right].
\label{eq:(3.10)}
\end{equation}
$U_{2\pi p/N}(\mathcal{M}_3)$ is topological by construction and it can be seen
that $U_{2\pi p/N}(\mathcal{M}_3)$ is invariant under the lattice gauge
transformation~\eqref{eq:(2.5)}~\cite{Honda:2024yte}. Thus, this generates a
physical symmetry transformation. In~Ref.~\cite{Honda:2024yte}, fusion rules
among the symmetry operators are computed by employing the lattice
representation~\eqref{eq:(3.8)}. It can also be seen that, from the
expression~\eqref{eq:(3.9)}, $U_{2\pi p/N}(\mathcal{M}_3)=0$
if~$\delta z\bmod N\neq0$ in~$H^2(\mathcal{M}_3;\mathbb{Z}_N)$, showing the
non-invertible nature of~$U_{2\pi p/N}(\mathcal{M}_3)$~\cite{Honda:2024yte}.

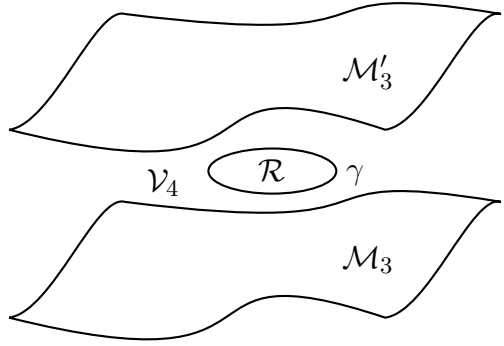
\begin{figure}[htbp]
\centering
\begin{tikzpicture}[scale=0.5]
\draw[thick](7,3.9,0) circle(1.7 and 0.6);
  \draw[thick] (0,5,0) ..controls (8,3,0) and (4,7,0).. (10,5,0) ..controls (10,4,-3) and (10,6,-6).. (10,5,-8) .. controls (4,6,-8) and (8,4,-8) .. (0,5,-8) ..controls (0,6,-6) and (0,4,-3).. (0,5,0);
  \node at (8,5,-4) {$\mathcal{M}_3'$};
  \draw[thick] (0,0,0) ..controls (8,-2,0) and (4,2,0).. (10,0,0) ..controls (10,-1,-3) and (10,1,-6).. (10,0,-8) .. controls (4,1,-8) and (8,-1,-8) .. (0,0,-8) ..controls (0,1,-6) and (0,-1,-3).. (0,0,0);
  \node at (8,0,-4) {$\mathcal{M}_3$};
  \node at (4.1,3.6,0) {$\mathcal{V}_4$};
  \node at (9.2,3.8,0) {$\gamma$};
  \node at (6.98,3.93,0) {$\mathcal{R}$};
\end{tikzpicture}
\caption{Schematic view of the configuration of the 3-surfaces $\mathcal{M}_3$
and~$\mathcal{M}_3'$ and the 't~Hooft line operator~$\gamma$.}
\label{fig:1}
\end{figure}

We now consider the situation depicted in~Fig.~\ref{fig:1}. That is, we
consider a 't~Hooft line~$\gamma=\partial(\mathcal{R})$ in a 4-volume
$\mathcal{V}_4$ and
$\partial(\mathcal{V}_4)=\mathcal{M}_3'\cup(-\mathcal{M}_3)$. We assume that
$\gamma$ is \emph{not\/} contained in~$\partial(\mathcal{V}_4)
=\mathcal{M}_3'\cup(-\mathcal{M}_3)$.
For this situation, the integration over the auxiliary field~$\Tilde{a}$
enforces
\begin{equation}
   \delta z=\mathrm{q}\delta_3[\gamma]
   =\mathrm{q}\delta\delta_2[\mathcal{R}],
\label{eq:(3.11)}
\end{equation}
where $\delta_2[\mathcal{R}]$ is the delta function 2-cochain
on~$\mathcal{R}$. Therefore, the 2-cochain $z-\mathrm{q}\delta_2[\mathcal{R}]$
is coclosed in~$\mathbb{Z}$. We then note that a deformation of the 3-surface
$\mathcal{M}_3'$ can be regarded as an attachment of a 4-ball $\mathcal{B}_4$
to~$\mathcal{M}_3'$. Since the second cohomology group is trivial for a 4-ball,
$H^2(\mathcal{B}_4,\mathbb{Z})=0$, there exists a
1-cochain~$\nu\in C^1(\mathcal{B}_4,\mathbb{Z})$ such that
\begin{equation}
   z=\delta\nu+\mathrm{q}\delta_2[\mathcal{R}].
\label{eq:(3.12)}
\end{equation}
Since such a deformation of the 3-surface can be repeated until
$\mathcal{M}_3'=\mathcal{M}_3$ (especially on a finite periodic lattice), this
argument shows that there exists a
1-cochain~$\nu\in C^1(\mathcal{V}_4,\mathbb{Z})$ such that
Eq.~\eqref{eq:(3.12)} holds for~$\mathcal{V}_4$~\cite{Honda:2024yte}.

Going back to the anomalous WT identity~\eqref{eq:(3.6)}, let us compute the
exponent on the left-hand side of~Eq.~\eqref{eq:(3.6)} by using
Eqs.~\eqref{eq:(2.14)} and~\eqref{eq:(3.12)}. After some calculation, we find
\begin{equation}
   f\cup f
   =\delta\left(a\cup f+2\pi z\cup a+4\pi^2\delta\nu\cup\nu\right)
   +4\pi\mathrm{q}f\cup\delta_2[\mathcal{R}]
   -2\pi\mathrm{q}f\cup_1\delta_3[\gamma].
\label{eq:(3.13)}
\end{equation}
Here, we have neglected $\delta_3[\gamma]$ and~$\delta_2[\mathcal{R}]$
inside $\delta(\dotsb)$, because under the integration over~$\mathcal{V}_4$,
those do not contribute. Although $\nu$ in~Eq.~\eqref{eq:(3.12)} is ambiguous
by a 1-cocycle~$c\in\mathbb{Z}$, $\delta c=0$, as $\nu\to\nu+c$, this ambiguity
does not affect Eq.~\eqref{eq:(3.13)} because
$\delta(\delta\nu\cup c)=\delta^2(\nu\cup c)=0$. Then,
\begin{align}
   -\frac{i\alpha}{2}\frac{\mathrm{e}^2}{4\pi^2}
   \sum_{\text{hypercube}\in\mathcal{V}_4}f\cup f
   &=-\frac{i\alpha}{2}\frac{\mathrm{e}^2}{4\pi^2}
   \sum_{\text{cube}\in\mathcal{M}_3'\cup(-\mathcal{M}_3)}
   \left(a\cup f+2\pi z\cup a+4\pi^2\delta\nu\cup\nu\right)
\notag\\
   &\qquad{}
   -\frac{i\mathrm{q}\mathrm{e}^2\alpha}{2\pi}
   \sum_{\text{hypercube}\in\mathcal{V}_4}
   \left(f\cup\delta_2[\mathcal{R}]
   -\frac{1}{2}f\cup_1\delta_3[\gamma]\right).
\label{eq:(3.14)}
\end{align}
The last term precisely cancels the exponential on the right-hand side
of~Eq.~\eqref{eq:(3.6)} and thus the anomalous WT identity yields,
for the rotation angle~\eqref{eq:(3.7)},
\begin{align}
   &\left\langle
   \exp\left\{
   \frac{i\pi p}{N}\sum_{\text{cube}\in\mathcal{M}_3'\cup(-\mathcal{M}_3)}
   \left[\star j_5
   -\frac{\mathrm{e}^2}{4\pi^2}
   \left(a\cup f+2\pi z\cup a+4\pi^2\delta\nu\cup\nu\right)
   \right]
   \right\}T_{\mathrm{q}}(\gamma)\dotsb
   \right\rangle_{\mathrm{B}}
\notag\\
   &=
   \left\langle
   T_{\mathrm{q}}(\gamma)\dotsb
   \right\rangle_{\mathrm{B}}.
\label{eq:(3.15)}
\end{align}
Since the right-hand side of~Eq.~\eqref{eq:(3.13)} and thus its integration
over~$\mathcal{V}_4$, the right-hand side of~Eq.~\eqref{eq:(3.14)}, are
independent from the ambiguity of~$\nu$ under~$\nu\to\nu+c$, the left-hand side
of this expression is also free from this ambiguity in~$\nu$ (the last term
of~Eq.~\eqref{eq:(3.14)} is obviously independent of the choice of~$\nu$).

On the other hand, in the above situation, from~Eq.~(C14)
of~Ref.~\cite{Honda:2024yte},
\begin{align}
   \mathcal{Z}_{\mathcal{M}_3'}[z]
   &=\exp\left(
   -\frac{i\pi p\mathrm{e}^2}{N}
   \sum_{\mathrm{cube}\in\mathcal{M}_3'}\delta\nu\cup\nu
   \right)
   \mathcal{Z}_{\mathcal{M}_3'}[0]
\notag\\
   &=\exp\left(
   -\frac{i\pi p\mathrm{e}^2}{N}
   \sum_{\mathrm{cube}\in\mathcal{M}_3'}\delta\nu\cup\nu
   \right)
   \mathcal{Z}_{\mathcal{M}_3}[0]
\notag\\
  &=\exp\left[
   -\frac{i\pi p\mathrm{e}^2}{N}
   \left(
   \sum_{\mathrm{cube}\in\mathcal{M}_3'}\delta\nu\cup\nu
   -\sum_{\mathrm{cube}\in\mathcal{M}_3}\delta\nu\cup\nu
   \right)
   \right]
   \mathcal{Z}_{\mathcal{M}_3}[z],
\label{eq:(3.16)}
\end{align}
where $\mathcal{Z}_{\mathcal{M}_3}[0]=N^{b_2-1}$ ($b_2$ is the second Betti
number of~$\mathcal{M}_3$). Here, we have noted
$\mathcal{Z}_{\mathcal{M}_3'}[0]=\mathcal{Z}_{\mathcal{M}_3}[0]$ because
$\mathcal{M}_3'$ and~$\mathcal{M}_3$ are homologically equivalent,
$\partial(\mathcal{V}_4)=\mathcal{M}_3'\cup(-\mathcal{M}_3)$. Note that this
relation holds irrespective of the existence of the 't~Hooft line
in~$\mathcal{V}_4$.\footnote{Also, this relation should hold for any choice of
TQFT to represent the symmetry operator because, when $\delta z=0$
within~$\mathcal{V}_4$, from~Eq.~\eqref{eq:(3.11)}, we should
have~\cite{Honda:2024yte},
\begin{align}
   \mathcal{Z}_{\mathcal{M}_3'}[z]
   &=\exp\left[
   -\frac{i\pi p\mathrm{e}^2}{N}
   \sum_{\mathrm{hypercube}\in\mathcal{V}_4}z\cup z
   \right]
   \mathcal{Z}_{\mathcal{M}_3}[z]
\notag\\
   &=\exp\left[
   -\frac{i\pi p\mathrm{e}^2}{N}
   \left(
   \sum_{\mathrm{cube}\in\mathcal{M}_3'}\delta\nu\cup\nu
   -\sum_{\mathrm{cube}\in\mathcal{M}_3}\delta\nu\cup\nu
   \right)
   \right]
   \mathcal{Z}_{\mathcal{M}_3}[z].
\label{eq:(3.17)}
\end{align}
}
Eqs.~\eqref{eq:(3.16)} and~\eqref{eq:(3.8)} show that the relation
\begin{align}
   &U_{2\pi p/N}(\mathcal{M}_3')
\notag\\
   &=\exp\left\{
   \frac{i\pi p}{N}\sum_{\text{cube}\in\mathcal{M}_3'\cup(-\mathcal{M}_3)}
   \left[\star j_5
   -\frac{\mathrm{e}^2}{4\pi^2}
   \left(a\cup f+2\pi z\cup a+4\pi^2\delta\nu\cup\nu\right)
   \right]
   \right\}
   U_{2\pi p/N}(\mathcal{M}_3).
\label{eq:(3.18)}
\end{align}

Finally, we substitute $\dotsb$ in~Eq.~\eqref{eq:(3.15)}
by~$U_{2\pi p/N}(\mathcal{M}_3)$. We can do this because it can be seen that
$U_{2\pi p/N}(\mathcal{M}_3)$~\eqref{eq:(3.8)} is invariant under the shift of
the scalar field within~$\mathcal{V}_4$. Then, using~Eq.~\eqref{eq:(3.18)}, we
have
\begin{equation}
   \left\langle
   U_{2\pi p/N}(\mathcal{M}_3')T_{\mathrm{q}}(\gamma)
   \right\rangle_{\mathrm{B}}
   =
   \left\langle
   U_{2\pi p/N}(\mathcal{M}_3)
   T_{\mathrm{q}}(\gamma)
   \right\rangle_{\mathrm{B}}.
\label{eq:(3.19)}
\end{equation}
This is our main result. Schematically speaking, when the symmetry operator
sweeps out a 't~Hooft line operator as~Fig.~\ref{fig:1}, it does not leave
any factor. Note that Eq.~\eqref{eq:(3.19)} has been obtained without
specifying what happens when the symmetry operator collides with the 't~Hooft
line operator. Therefore, Eq.~\eqref{eq:(3.19)} holds irrespective of possible
modifications of the symmetry operator $U_{2\pi p/N}(\mathcal{M}_3)$ for the
situation~$\gamma\subset\mathcal{M}_3$. This our result appears inequivalent
with the phenomenon concluded in~Refs.~\cite{Choi:2022jqy,Cordova:2022ieu} in
the continuum theory.

For the axion string operator~\eqref{eq:(2.27)}, since its dressing factor
does not contain the scalar field and the change of the
action~\eqref{eq:(2.24)} under the axial transformation
$\phi(x)\to\phi(x)+\alpha(x)$ does not contain~$\delta\ell$, we infer that
the axion string does not receive any effect under the sweep by the symmetry
operator. It might be interesting to consider the situation in which the
't~Hooft line and the axion string coexist; the dressing factors
in~Eqs.~\eqref{eq:(2.26)} and~\eqref{eq:(2.27)} should then be further modified
for the gauge invariance.

\section*{Acknowledgments}
We would like to thank
Motokazu Abe,
Okuto Morikawa,
and
Yuya Tanizaki
for discussions which motivated the present work.
This work was partially supported by Japan Society for the Promotion of Science
(JSPS) Grant-in-Aid for Scientific Research Grant Number JP23K03418~(H.S.).

\section*{Note added}
In~Ref.~\cite{Honda:2024xmk}, we present an alternative and much simpler
derivation of the conclusion of this paper.

\appendix

\section{The half-space gauging leads to the same conclusion}
\label{sec:A}
In this appendix, we demonstrate that the half-space gauging of the magnetic
$\mathbb{Z}_N$ 1-form symmetry, which was employed
in~Refs.~\cite{Choi:2022jqy,Cordova:2022ieu} to study the action of the
symmetry operator on the 't~Hooft line operator, when formulated in an
appropriate lattice framework, leads to the same conclusion as that in the main
text. That is, the sweep of the symmetry operator over the 't~Hooft line does
not leave any trace.

For this, for an integer~$T\geq0$, we introduce a half-space lattice (we
consider an infinite lattice for simplicity) by
\begin{equation}
   \Gamma^+(T)
   :=\left\{x\in\mathbb{Z}^4\mathrel{}\middle|\mathrel{}x_0\geq T\right\}.
\label{eq:(A1)}
\end{equation}
We assume that the 't Hooft line operator~$T_{\mathrm{q}}(\gamma)$ along a
loop~$\gamma$ lies on the 3-plane~$x_0=T_\gamma>0$.

Now, corresponding to~Eq.~(2.18) of~Ref.~\cite{Choi:2022jqy}, the half-spacing
gauging of the magnetic $\mathbb{Z}_N$ 1-form symmetry in the
region~$\Gamma^+(T)$ is implemented by, to our lattice system $e^{-S}$, where
$S$ is given by~Eq.~\eqref{eq:(2.24)}, supplementing the partition
function,\footnote{In this appendix, we denote the hypercube, cube, plaquette,
and link simply by $h$, $c$, $p$, and~$l$, respectively.}
\begin{align}
   \int\mathrm{D}[\xi]\mathrm{D}[\eta]\,
   \exp
   \left\{
   -\frac{i\pi p\mathrm{e}^2}{N}\sum_{h\in\Gamma^+(T)}
   \left[
   P_2(\xi-z)-P_2(z)
   +2\eta\cup\delta\xi
   \right]
   \right\}.
\label{eq:(A2)}
\end{align}
Here, $\xi$ and $\eta$ are 2-cochain and 1-cochain on~$\Gamma^+(T)$,
respectively\footnote{$(2\pi p/N)\xi$ and~$(2\pi/N)\eta$ correspond to $b$
and~$c$ in~Ref.~\cite{Choi:2022jqy} (where $\mathrm{e}=1$), respectively; we
use these notations to avoid possible confusion with the fields
in~Eq.~\eqref{eq:(3.9)}.} and $P_2(\xi)$ denotes the Pontryagin square;
\begin{equation}
   P_2(\xi):=\xi\cup\xi+\xi\cup_1\delta\xi.
\label{eq:(A3)}
\end{equation}
The integration measures in~Eq.~\eqref{eq:(A2)} are given by
\begin{equation}
   \mathrm{D}[\xi]:=
   \sum_{p\in\Gamma^+(T)}\left[
   \sum_{\xi_{\mu\nu}(x)=0}^{N-1}
   \right],\qquad
   \mathrm{D}[\eta]:=
   \sum_{l\in\Gamma^+(T)}\left[
   \frac{1}{N}\sum_{\eta_\mu(x)=0}^{N-1}
   \right],
\label{eq:(A4)}
\end{equation}
and $\xi$ is supposed to obey the Dirichlet boundary condition at the
boundary~$\partial\Gamma^+(T)$;
\begin{equation}
   \xi_{12,23,31}(x_0=T)=0\bmod N.
\label{eq:(A5)}
\end{equation}
The 2-cochain~$\xi$ represents a gauge field, which couples linearly to the
2-cochain~$z$; $z$ is a magnetic ``current'' because it conserves when the
Bianchi identity~$\delta z=0$~\eqref{eq:(2.7)} holds. The 1-cochain~$\eta$
is a Lagrange multiplier and its integration produces the constraint,
\begin{equation}
   \delta\xi=0\bmod N.
\label{eq:(A6)}
\end{equation}

First, let us examine the invariance of the action in~Eq.~\eqref{eq:(A2)} under
the magnetic $\mathbb{Z}_N$ 1-form gauge transformation
\begin{equation}
   \xi\to\xi+\delta\rho,\qquad
   \eta\to\eta-\rho,\qquad
   z\to z,
\label{eq:(A7)}
\end{equation}
where $\rho\in\mathbb{Z}_N$. Under these, we see that the action changes into
\begin{align}
   &P_2(\xi-z)-P_2(z)+2\eta\cup\delta\xi
\notag\\
   &\to
   P_2(\xi-z)-P_2(z)+2\eta\cup\delta\xi
\notag\\
   &\qquad{}
   -2\rho\cup\delta z
   +\delta\left[
   2\rho\cup(\xi-z)
   +\delta\rho\cup_1(\xi-z)
   +\rho\cup\delta\rho
   \right].
\label{eq:(A8)}
\end{align}
The term that is proportional to~$\delta z$ in this expression may be removed
by further transforming the dual $U(1)$ gauge potential~$\Tilde{a}$
in~Eq.~\eqref{eq:(2.24)} as
\begin{equation}
   \Tilde{a}\to\Tilde{a}+\frac{2\pi p\mathrm{e}^2}{N}\rho.
\label{eq:(A9)}
\end{equation}
We further assume that
\begin{equation}
   \rho_{1,2,3}(x_0=T)=0\bmod N
\label{eq:(A10)}
\end{equation}
for the boundary condition~\eqref{eq:(A5)} to be invariant under the gauge
transformation. Then, it can be seen that the surface terms in $\delta[\dotsb]$
in~Eq.~\eqref{eq:(A8)} at~$x_0=T$ vanish mod~$N$. Therefore, the system is
invariant under the 1-form gauge transformation defined by~Eqs.~\eqref{eq:(A7)}
and~\eqref{eq:(A9)}. We note that, once the coupling of the
form~$2\eta\cup\delta\xi$ in~Eq.~\eqref{eq:(A2)} to impose the
constraint~\eqref{eq:(A6)} is adopted, then the usage of the Pontryagin
square~$P_2(\xi-z)$, instead of the simple square~$(\xi-z)\cup(\xi-z)$, is
crucial for the above gauge invariance.

Next, for the gauging to produce a 3D TQFT on the
boundary~$\partial\Gamma^+(T)$, the partition function~\eqref{eq:(A2)} should
be invariant under the 1-form gauge transformation of~$z$,
\begin{equation}
   z\to z+\delta\omega,
\label{eq:(A11)}
\end{equation}
where $\omega\in\mathbb{Z}$. That is, the functional integration over
$\xi$ and~$\eta$ in~Eq.~\eqref{eq:(A2)} should reproduce the anomaly being
identical to the anomaly produced by the ``SPT action'',
\begin{equation}
   \exp\left[-\frac{i\pi p\mathrm{e}^2}{N}
   \sum_{h\in\Gamma^+(T)}P_2(z)\right].
\label{eq:(A12)}
\end{equation}
Noting the relation
\begin{equation}
   P_2(\xi-z)-P_2(z)
   =P_2(\xi)-2\xi\cup z
   -\delta(\xi\cup_1z)+\delta\xi\cup_1z-z\cup_1\delta\xi,
\label{eq:(A13)}
\end{equation}
we see that, under~Eq.~\eqref{eq:(A11)},
\begin{align}
   &P_2(\xi-z)-P_2(z)
\notag\\
   &\to P_2(\xi-z)-P_2(z)
   +2\delta\xi\cup\omega
   +\delta\xi\cup_1\delta\omega
   -\delta\omega\cup_1\delta\xi
   -\delta\left(
   2\xi\cup\omega
   +\xi\cup_1\delta\omega\right).
\label{eq:(A14)}
\end{align}
Under the constraint~\eqref{eq:(A6)} and the boundary
condition~\eqref{eq:(A5)}, we see that the partition function~\eqref{eq:(A2)}
is invariant under~Eq.~\eqref{eq:(A10)}. This shows that the functional
integral over $\xi$ and~$\eta$ in~Eq.~\eqref{eq:(A2)} gives the same anomaly
as~Eq.~\eqref{eq:(A12)}.

This is not quite enough, however, to ensure the existence of a 3D TQFT on the
boundary~$\partial\Gamma^+(T)$. For this, the anomaly arising
from~Eq.~\eqref{eq:(A12)} should localize on the
boundary~$\partial\Gamma^+(T)$. Under~Eq.~\eqref{eq:(A11)}.
\begin{equation}
   P_2(z)\to P_2(z)
   +2\omega\cup\delta z
   +\delta\left(
   2\omega\cup z+\delta\omega\cup_1z+\omega\cup\delta\omega
   \right).
\label{eq:(A15)}
\end{equation}
Therefore, the anomaly arising from~Eq.~\eqref{eq:(A12)} is not localized at
the boundary~$\partial\Gamma^+(T)$, if the Bianchi identity is broken in the
bulk, $x_0>T$. This occurs if the 't~Hooft line lies within the gauged
region~$\Gamma^+(T)$. Recalling the representation~\eqref{eq:(3.12)}, we may
write
\begin{equation}
   z:=z_0+z',\qquad
   z_0:=\delta\nu,\qquad
   z':=\mathrm{q}\delta_2[\mathcal{R}].
\label{eq:(A16)}
\end{equation}
Then, the functional integral, which correctly produces a 3D TQFT on the
boundary~$\Gamma^+(T)$ is given by, instead of~Eq.~\eqref{eq:(A2)},
\begin{align}
   \int\mathrm{D}[\xi]\mathrm{D}[\eta]\,
   \exp
   \left\{
   -\frac{i\pi p\mathrm{e}^2}{N}\sum_{h\in\Gamma^+(T)}
   \left[
   P_2(\xi-z_0)-P_2(z_0)
   +2\eta\cup\delta\xi
   \right]
   \right\}.
\label{eq:(A17)}
\end{align}
This point is crucially important for our conclusion.

Now, we first consider the situation in which the 't~Hooft line
operator~$T_{\mathrm{q}}(\gamma)$ is outside the region~$\Gamma^+(T)$, that is,
$T_\gamma<T$. Then, we do the axial rotation of matter fields with the rotation
angle~$2\pi p/N$ in the region~$\Gamma^+(T)$. Considering the change
of~$e^{-S}$, we will have
\begin{align}
   &e^{-S}\exp\left(\frac{i\pi p}{N}
   \left\{
   \sum_{c\in\partial\Gamma^+(T)}
   \left[\star j_5-\frac{\mathrm{e}^2}{4\pi^2}
   \left(
   a\cup f+2\pi z\cup a+2\pi a\cup_1\delta z
   \right)\right]
   -\mathrm{e}^2\sum_{h\in\Gamma^+(T)}P_2(z)
   \right\}\right)
\notag\\
   &\qquad{}
   \times
   \int\mathrm{D}[\xi]\mathrm{D}[\eta]\,
   \exp
   \left\{
   -\frac{i\pi p\mathrm{e}^2}{N}\sum_{h\in\Gamma^+(T)}
   \left[
   P_2(\xi-z)-P_2(z)+2\eta\cup\delta\xi
   \right]
   \right\}
   T_{\mathrm{q}}(\gamma).
\label{eq:(A18)}
\end{align}
This change of~$e^{-S}$ corresponds to the lattice axial anomaly of the form
$\delta\star j_5=[\mathrm{e}^2/(4\pi^2)]
[f\cup f-2\pi a\cup\delta z+2\pi\delta z\cup a+2\pi\delta(a\cup_1\delta z)
+4\pi^2z\cup_1\delta z]$; this is same as the anomaly arising from the shift of
the axion in the main text (Eq.~\eqref{eq:(3.2)}). Note that in this expression
of the anomaly, the terms that are square in~$z$ are contained in the form of
the Pontryagin square~$P_2(z)$ instead of~$z\cup z$. This form of the anomaly
(when $\delta z\neq0$) is consistent with the structure of the partition
function~\eqref{eq:(A2)} and the usage of~$P_2(z)$ in~Eq.~\eqref{eq:(A2)} is,
as elucidated above, closely linked to the magnetic gauge invariance.
Therefore, the form of the lattice axial anomaly in~Eq.~\eqref{eq:(A18)} is
quite natural irrespective of the underlying lattice action.

Equation~\eqref{eq:(A18)} is the expression of the expectation value of the
't~Hooft line operator~$T_{\mathrm{q}}(\gamma)$, when $T_{\mathrm{q}}(\gamma)$
lies outside the region~$\Gamma^+(T)$. Under the integration over~$\Tilde{a}$
in~Eq.~\eqref{eq:(2.2)}, therefore, $z$ in~Eq.~\eqref{eq:(A18)} fulfills the
Bianchi identity~$\delta z=0$ and we can set $z\to z_0$ in the expression.

Next, we enlarge the region~$\Gamma^+(T)$ to~$\Gamma^+(0)$ so that the 't~Hooft
line operator~$T_{\mathrm{q}}(\gamma)$ at~$x_0=T_\gamma>0$ is inside the gauged
region. This is achieved by further multiplying the factor
\begin{align}
   \int\mathrm{D}[\xi]\mathrm{D}[\eta]\,
   \exp
   \left\{
   -\frac{i\pi p\mathrm{e}^2}{N}\sum_{h\in\Gamma^+(0)-\Gamma^+(T)}
   \left[
   P_2(\xi-z)-P_2(z)
   +2\eta\cup\delta\xi
   \right]
   \right\},
\label{eq:(A19)}
\end{align}
to Eq.~\eqref{eq:(A18)}. Here, we cannot use $P_2(\xi-z_0)-P_2(z_0)$ instead
of~$P_2(\xi-z)-P_2(z)$, because, in the next step of the axial rotation
in~$\Gamma^+(0)-\Gamma^+(T)$, the axial anomaly produces the factor~$P_2(z)$;
we want to cancel this $P_2(z)$ by the $P_2(z)$ in~Eq.~\eqref{eq:(A19)} to make
the defect topological.

For the gauging of the magnetic 1-form symmetry in the enlarged
region~$\Gamma^+(0)$, however, this is not sufficient and we also have to
modify~$T_{\mathrm{q}}(\gamma)$ as
\begin{align}
   T_{\mathrm{q}}(\gamma)\to
   T_{\mathrm{q}}(\gamma)\exp\left(-i\mathrm{q}\frac{2\pi p\mathrm{e}^2}{N}
   \sum_{p\in\mathcal{R}}\xi\right),
\label{eq:(A20)}
\end{align}
where $\partial\mathcal{R}=\gamma$, for the invariance
under~Eqs.~\eqref{eq:(A7)} and~\eqref{eq:(A9)} (recall Eq.~\eqref{eq:(2.8)}).
The factor in~Eq.~\eqref{eq:(A20)} corresponds to the surface operator
associated with the line operator considered in~Ref.~\cite{Choi:2022jqy}
(Eq.~(2.28) there). This factor required for the magnetic 1-form gauge
invariance may also be termed dressing.

Then, we again apply the axial rotation on matter fields, but this time only in
the region~$\Gamma^+(0)-\Gamma^+(T)$. Then, as~Eq.~\eqref{eq:(A18)}, we will
have
\begin{align}
   &e^{-S}\exp\left(\frac{i\pi p}{N}
   \left\{
   \sum_{c\in\partial\Gamma^+(0)}
   \left[\star j_5-\frac{\mathrm{e}^2}{4\pi^2}
   \left(
   a\cup f+2\pi z\cup a+2\pi a\cup_1\delta z
   \right)\right]
   -\mathrm{e}^2\sum_{h\in\Gamma^+(0)}P_2(z)
   \right\}\right)
\notag\\
   &\qquad{}
   \times
   \int\mathrm{D}[\xi]\mathrm{D}[\eta]\,
   \exp
   \left\{
   -\frac{i\pi p\mathrm{e}^2}{N}\sum_{h\in\Gamma^+(0)}
   \left[
   P_2(\xi-z)-P_2(z)+2\eta\cup\delta\xi
   \right]
   \right\}
\notag\\
   &\qquad\qquad{}
   \times
   T_{\mathrm{q}}(\gamma)
   \exp\left(i\mathrm{q}\frac{2\pi p\mathrm{e}^2}{N}
   \sum_{p\in\mathcal{R}}z\right)
   \exp\left(-i\mathrm{q}\frac{2\pi p\mathrm{e}^2}{N}
   \sum_{p\in\mathcal{R}}\xi\right).
\label{eq:(A21)}
\end{align}
Here, we have assumed that the 't~Hooft line~$T_{\mathrm{q}}(\gamma)$
produces the factor
\begin{equation}
   \exp\left(i\mathrm{q}\frac{2\pi p\mathrm{e}^2}{N}
   \sum_{p\in\mathcal{R}}z\right)
\label{eq:(A22)}
\end{equation}
under the axial rotation. This is actually the case for the QED axion in the
main text (see~Eq.~\eqref{eq:(3.4)}) but we expect that \emph{this is generally
true}. The reason is that this factor is required for the invariance under the
$\mathbb{Z}$ 1-form gauge transformation~\eqref{eq:(2.18)}. In fact, since the
combination in the exponent in the first line of~Eq.~\eqref{eq:(A21)} is
expressed as
\begin{equation}
   -\frac{i\pi p\mathrm{e}^2}{4\pi^2N}\sum_{h\in\Gamma^+(0)}
   \left[f\cup f-2\pi a\cup\delta z+2\pi\delta z\cup a
   +2\pi\delta(a\cup_1\delta z)+4\pi^2z\cup_1\delta z\right],
\label{eq:(A23)}
\end{equation}
under~Eq.~\eqref{eq:(2.18)}, this changes by
\begin{equation}
   \frac{2\pi ip\mathrm{e}^2}{N}\sum_{h\in\Gamma^+(0)}
   \left[m\delta z-\frac{1}{2}\delta\left(m\cup_1\delta z\right)\right]
   =\frac{2\pi ip\mathrm{e}^2}{N}\mathrm{q}\sum_{p\in\mathcal{R}}m,
\label{eq:(A24)}
\end{equation}
where we have used~Eq.~\eqref{eq:(A16)} and assumed that $T_\gamma\gg1$ (or
we may assume that $m_{1,2,3}(x_0=0)=0$). Since the second line
of~Eq.~\eqref{eq:(A21)} is invariant under~$z\to z-\delta m$
(recall~Eq.~\eqref{eq:(A11)}), we see that the change of the
factor~\eqref{eq:(A22)} under~$z\to z-\delta m$ makes Eq.~\eqref{eq:(A21)}
$\mathbb{Z}$ 1-form gauge invariant. Thus, if the lattice formulation is
sound,\footnote{The $\mathbb{Z}$ 1-form gauge invariance
under~Eq.~\eqref{eq:(2.18)} is a basic principle of the modified Villain
formulation that we have employed in this paper. In the compact $U(1)$ lattice
gauge theory, where $f=\delta a+2\pi z$, the invariance
under~Eq.~\eqref{eq:(2.18)} is simply a parametrization redundancy and thus
should automatically holds.} we expect that the factor~\eqref{eq:(A22)} always
emerges associated with the axial rotation.

For the representation~\eqref{eq:(A16)}, we find
\begin{align}
   P_2(\xi-z)
   &=P_2(\xi-z_0)-2(\xi-z_0)\cup z'+z'\cup z'+z'\cup_1\delta z'
\notag\\
   &\qquad{}
   -\delta[(\xi-z_0)\cup_1z']
   +\delta(\xi-z_0)\cup_1 z'-z'\cup_1\delta(\xi-z_0).
\label{eq:(A25)}
\end{align}
Using $\delta z_0=0$, $\delta\xi=0\bmod N$~\eqref{eq:(A6)}, and the fact that
$z'\neq0$ only for~$x_0>0$, we may further simplify this as
\begin{equation}
   P_2(\xi-z)
   =P_2(\xi-z_0)-2\mathrm{q}(\xi-z_0)\cup\delta_2[\mathcal{R}],
\label{eq:(A26)}
\end{equation}
because $z'\cup z'=0$ and~$z'\cup_1\delta z'=0$ by an explicit calculation
(see, e.g., Eq.~(A.9) of~Ref.~\cite{Jacobson:2023cmr}). The last term on the
right-hand side can be regarded as the Witten effect. Plugging this
into~Eq.~\eqref{eq:(A21)}, we finally arrive at
\begin{align}
   &e^{-S}\exp\left(\frac{i\pi p}{N}
   \left\{
   \sum_{c\in\partial\Gamma^+(0)}
   \left[\star j_5-\frac{\mathrm{e}^2}{4\pi^2}
   \left(
   a\cup f+2\pi z\cup a+2\pi a\cup_1\delta z
   \right)\right]
   -\mathrm{e}^2\sum_{h\in\Gamma^+(0)}P_2(z_0)
   \right\}\right)
\notag\\
   &\qquad{}
   \times
   \int\mathrm{D}[\xi]\mathrm{D}[\eta]\,
   \exp
   \left\{
   -\frac{i\pi p\mathrm{e}^2}{N}\sum_{h\in\Gamma^+(0)}
   \left[
   P_2(\xi-z_0)-P_2(z_0)+2\eta\cup\delta\xi
   \right]
   \right\}
   T_{\mathrm{q}}(\gamma).
\label{eq:(A27)}
\end{align}
Comparing this with~Eq.~\eqref{eq:(A18)}, where~$z=z_0$, and remembering that
the functional integral in~Eq.~\eqref{eq:(A17)} produces the desired
topological defect, we infer that the sweep of the symmetry operator over the
't~Hooft line operator does not leave any trace, the same conclusion as that in
the main text. It might be regarded as a consequence of a cancellation between
the dressing and the Witten effect.

Although the above demonstration holds, strictly speaking, only in the modified
Villain lattice formulation of the QED axion in the main text, in various
stages in the argument, we have indicated the general validity of our argument
on the basis of required symmetries. It is thus interesting to reinforce our
argument from various other perspectives.

% can use a bibliography generated by BibTeX as a .bbl file
% BibTeX documentation can be easily obtained at:
% http://www.ctan.org/tex-archive/biblio/bibtex/contrib/doc/

% can use a bibliography generated by BibTeX as a .bbl file
% BibTeX documentation can be easily obtained at:
% http://www.ctan.org/tex-archive/biblio/bibtex/contrib/doc/

%\bibliographystyle{ptephy}
%\bibliography{sample}
%
% once the .bbl file has been generated then place the text in your article.

%% \vspace{0.2cm}
%% \noindent
%% For references, note how to include DOI information from examples below. 

%This is added by T. Yoneya (editor-in-chief) on 2020/07/09.

\let\doi\relax

%without this code before the command "\begin{thebibliography}{}" , an error will be %flagged. When the bibliography is provided as separate .bib file, then this code %should be placed above the commands "\bibliographystyle{}" and "\bibliography{}" %inside the main TeX file. 

%% \begin{thebibliography}{9}

%% \bibitem{1}
%% J. P.~Blaizot, and E.~Iancu, Phys. Rep. {\bf 359}, 355 (2002).
%% \doi{https://doi.org/10.1016/S0370-1573(01)00061-8}

%% \bibitem{2}
%% M.~Gyulassy, and L.~McLerran, Nucl.\ Phys.\  A {\bf 750}, 30 (2005). \\ \doi{https://doi.org/10.1016/j.nuclphysa.2004.10.034}

%% \bibitem{3}
%% S.~Aoki et al. [JLQCD Collaboration], Phys. Rev. D 72, 054510 (2005). \\
%% \doi{https://doi.org/10.1103/PhysRevD.72.05451}

%% \bibitem{4}
%% S.~Alekhin, A.~Djouadi, and S.~Moch, Phys. Lett. B 716, 214 (2012) [arXiv:1207.0980 [hep-ph]]. \doi{https://doi.org/10.1016/j.physletb.2012.08.024}

%% \end{thebibliography}

%\bibliographystyle{ptephy}
%\bibliography{anomaly}

\end{document}